\documentclass[manuscript]{aastex61}
 \usepackage{graphicx}
\begin{document}

\title{A multi-peak solar flare with a high turnover frequency of the gyrosynchrotron spectra from the loop-top source}

\author[0000-0002-6985-9863]{Zhao Wu}
\affil{School of Space Science and physics, Shandong University, Weihai, Shandong 264209, China; yaochen@sdu.edu.cn}
\affil{Laboratory for Electromagnetic Detection, Institute of Space Sciences, Shandong University, Weihai, Shandong 264209, China}
%\affil{Shandong Provincial Key Laboratory of Optical Astronomy and Solar-Terrestrial Environment, and Institute of Space Sciences, Shandong University, Weihai, Shandong 264209, People¡¯s Republic of China}

\author[0000-0001-8644-8372]{Alexey Kuznetsov}
\affil{Institute of Solar-Terrestrial Physics, Irkutsk 664033, Russia}

\author[0000-0002-1107-7420]{Sergey Anfinogentov}
\affil{Institute of Solar-Terrestrial Physics, Irkutsk 664033, Russia}

\author{Victor Melnikov}
\affil{Central Astronomical Observatory at Pulkovo of the Russian Academy of Sciences, 196140 St Petersburg, Russia}

\author{Robert Sych}
\affil{Institute of Solar-Terrestrial Physics, Irkutsk 664033, Russia}

\author{Bing Wang}
\affil{Laboratory for Electromagnetic Detection, Institute of Space Sciences, Shandong University, Weihai, Shandong 264209, China}

\author[0000-0002-2734-8969]{Ruisheng Zheng}
\affil{School of Space Science and physics, Shandong University, Weihai, Shandong 264209, China; yaochen@sdu.edu.cn}

\author[0000-0003-1034-5857]{Xiangliang Kong}
\affil{School of Space Science and physics, Shandong University, Weihai, Shandong 264209, China; yaochen@sdu.edu.cn}

\author{Baolin Tan}
\affil{CAS Key Laboratory of Solar Activity, National Astronomical Observatories, Chinese Academy of Sciences, Beijing 100101, China}

\author{Zongjun Ning}
\affil{Key Laboratory of Dark Matter and Space Science, Purple Mountain Observatory, Chinese Academy of Sciences, Nanjing 210023, China}

\author[0000-0001-6449-8838]{Yao Chen}
\affil{School of Space Science and physics, Shandong University, Weihai, Shandong 264209, China; yaochen@sdu.edu.cn}
\affil{Laboratory for Electromagnetic Detection, Institute of Space Sciences, Shandong University, Weihai, Shandong 264209, China}
\affil{Institute of Frontier and Interdisciplinary Science, Shandong University, Qingdao, Shandong 266237, China}

\begin{abstract}
{The origin of multiple peaks in lightcurves of various wavelengths remains illusive during flares. Here we discuss the flare of SOL2023-05-09T03:54M6.5 with six flux peaks as recorded by a tandem of new microwave and Hard X-ray instruments. According to its microwave spectra, the flare represents a high-turnover frequency ($>$15 GHz) event. The rather-complete microwave and HXR spectral coverage provides a rare opportunity to uncover the origin of such event together with simultaneous EUV images. We concluded that (1) the microwave sources originates around the top section of the flaring loops with a trend of source spatial dispersion with frequency; (2) the visible movement of the microwave source from peak to peak originates from the process of new flaring loops appearing sequentially along the magnetic neutral line; 3) the optically-thin microwave spectra are hard with the indices ($\alpha_{tn}$) varying from~$\sim$$-1.2$ to $-0.4$, and the turnover frequency always exceeds 15~GHz; 4) higher turnover/peak frequency corresponds to stronger peak intensity and harder optically-thin spectra. Using the Fokker-Planck and GX simulator codes we obtained a good fit to the observed microwave spectra and spatial distribution of the sources at all peaks, if assuming the radiating energetic electrons have the same spatial distribution and single-power-law spectra but with the number density varying in a range of $\sim30\%$. We conclude that the particle acceleration in this flare happens in a compact region nearing the looptop. These results provide new constraints on the acceleration of energetic electrons and the underlying flare intermittent reconnection process.
}
\end{abstract}

\section{Introduction}
Solar flares often manifest multiple peaks in their light curves of microwave and hard X-ray (HXR) emissions. The underlying physics for these peaks remains ambiguous, and may vary from event to event. Such events are often classified as quasi-periodic pulsations (QPP) events, if the peaks present obvious periodicity \citep[e.g.][]{Nakariakov2009,Zimovets2021,Li2022}. Possible mechanisms include the modulation of emission and energy release by magnetohydrodynamic (MHD) waves \citep{Reznikova2010,Mclaughlin2018}, the self oscillation caused by steady inflow towards the reconnection site \citep{Nakariakov2010}, and intermittent reconnection \citep[e.g.][]{Li2006,Ofman2011,Liu2013,Kim2013,Wu2016}. \citet{Ning2018} investigated the flare of SOL2016-07-24T06:20 with double HXR peaks. They concluded the peaks are due to a two-stage energy release process, with the first peak being non-thermal due to energetic electrons accelerated via the loop-loop reconnection, and the second peak being thermally-dominated due to direct heating through the loop-loop reconnection at a relatively high altitude. This means the peaks may have distinct physical origin. These studies reveal the complex physical nature underlying the radiation peaks of solar flares.

Analysis combining the microwave and HXR data is essential to probe the physics of flares \citep[e.g.][]{White2011,Chen2016,Chen2017,Gary2018,Nindos2020,Krucker2020}. A tandem of instruments have been developed to do this, including: 1) the Siberian RadioHeliograph \citep[SRH,][]{Altyntsev2020} with three separated sub-arrays in bands of 3-6, 6-12, and 12-24 GHz, with test observation starting since March 2021 and regular observation since December 2023; 2) the Chashan Broadband Solar millimeter spectrometer \citep[CBS,][]{Shang2022,Shang2023}, recording the dynamic spectra from 35 to 40~GHz; 3) the Hard X-Ray Imager \citep[HXI,][]{Su2019} on board the Advanced Space-based Solar Observatory \citep[ASO-S,][]{Gan2023}, providing the HXR data from 10 to 300~keV; 4) the Nobeyama Radiopolarimeter \citep[NoRP,][]{Nakajima1985}, measuring the microwave flux density at 7 discrete frequencies from 1 to 80~GHz; and 5) Konus-Wind \citep[KW,][]{Lysenko2022}, recording the HXR flux in energy range of $\sim$18~keV-1.3~MeV.

With the CBS and NoRP data, \citet{Yan2023} reported similar spectral properties of flare peaks observed during the X2.2 flare on 2022 April 20. The event has a high turnover frequency ($\nu_{t}$) of its microwave gyrosynchrotron spectrum that extends from 20~GHz to $>$40~GHz during the impulsive stage. They reported a power-law dependence of the turnover flux $I_{t}$ on the turnover or peak frequency $\nu_{t}$, and identified the rapid-hardening-then-softening trend within the optically-thin regime of the gyrosynchrotron (GS) radiation. Usual GS spectra of flares present a reversed-V shape, with flux density peaking at $\nu_{t}$ that is $\sim$5-10~GHz for average events. Flares with a high turnover frequency ($>$15-20~GHz) are of particular interest since this indicates the abundance of mildly-relativistic electrons spiraling within a relatively strong magnetic field \citep[e.g.][]{White1992,Nignibeda2013,Song2016,Wu2019}. No microwave imaging data are available for the event. This limits further analysis of the radiation sources.

Here we focus on the M6.5 class flare on 2023 May 9 that has been well observed by the instruments mentioned above, with imaging data being available. During its impulsive stage the flare manifests 6 peaks in its microwave and HXR flux curves, all these peaks are characterized by high-turnover frequency. This provides a rare opportunity of further investigation of such event.

\section{Data and Event Overview}
\subsection{Instruments and Data}
The microwave images were observed by the low- ($3-6$~GHz) and middle-frequency ($6-12$~GHz) SRH arrays, with the time resolution of $\sim$3.5~seconds, spatial resolution of $15-30''$ and $12-24''$, and spectral resolution of 0.2 and 0.4~GHz. Note that the spatial resolution only means the smallest details that can be resolved, and the centroid position can be estimated with an error ($\sigma$) within a fraction of a beam width \citep[e.g.][]{Condon1997,Yu2024}
\begin{equation}
\sigma\approx\frac{\theta}{2\rm{SNR}\sqrt{\rm{ln}2}},
\end{equation}
where $\theta$ is the half-power beamwidth of the beam size, and SNR is the signal to noise ratio of the synthesized images (typically $\sim$1000 for SRH).

To minimize the instrumental shifts that may arise in the interferometric images, we further aligned the SRH microwave sources with the features visible in other spectral ranges (in particular, in magnetograms). We created a model with the magnetogram (at 03:48~UT) of the nearby non-flaring active region AR 13297, using the GX Simulator code \citep{Nita2023} and the technique described by \citet{Fleishman2021}. We derived position deviations (i.e., effectively, the differences between the observed and model source centroid positions, because the considered active region was only partially resolved with the SRH) for each frequency with cross-correlation of SRH observations and synthetic images of gyroresonance emission. We then applied the shifts required to remove these deviations (assuming them being constant with time) to all SRH images throughout the flare. The described procedure also allowed us to obtain a more accurate mutual alignment of the SRH images at different frequencies.

The following spectral data were used: 1) the NoRP radio flux densities at 1, 2, 3.75, 9.4, 17, and 35~GHz; 2) the integrated radio flux densities from the flaring region of the above SRH images; 3) the CBS radio flux densities from 35.25 to 39.75~GHz with steps of 500~MHz.

We analyzed HXR light curves recorded by HXI at discrete bands ($10-20$, $20-50$, $50-100$, and $100-300$~keV), and by the KW at $18-80$ and $80-328$~keV. We also used the Soft X-Ray (SXR) data recorded by the GOES satellites at $0.5-4.0$ and $1-8$~\AA{}.

\subsection{Event Overview}
\begin{figure}
\centering
\epsscale{.7}
\plotone{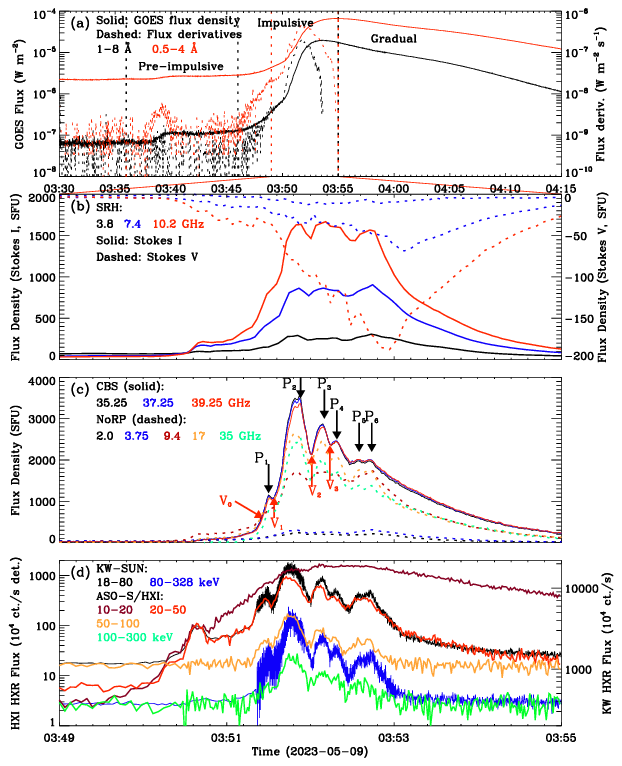}
\caption{Flux curves of the M6.5 flare observed on May 9 2023: panel (a) the SXR fluxes (solid) and their time derivatives (dashed) observed at 1-8~\AA{} (black) and 0.5-4~\AA{} (red) by GOES; panel (b) the integrated total (Stokes $I$, solid) and polarized (Stokes $V$, dashed) flux density observed by SRH at 3.8 (black), 7.4 (blue) and 10.2~GHz (red); panel (c) the flux density observed by CBS (solid) at 35.25 (black), 37.25 (blue), and 39.25 (red)~GHz, and by NoRP (dashed) at 2.0 (black), 3.75 (blue), 9.4 (red), 17 (orange), and 35 (green)~GHz; panel (d) the HXRs flux observed by HXI/ASO-S at 10-20 (brown), 20-50 (red), 50-100 (orange), and 100-300~keV (green), and by KW at 10-80 (black) and 80 - 328~keV (blue). Black arrows in panel (c) denote the local peaks of the flare, and red arrows denote the corresponding local valleys of the flux. }
 \label{fig1}
\end{figure}

The M6.5 flare on 2023 May 9 originated from the NOAA Active Region (AR) 13296 on the solar disk. According to the GOES SXR fluxes (solid lines in Figure~\ref{fig1}(a)), the eruption started at $\sim$03:36~UT, and peaked at $\sim$03:54~UT. The temporal profiles of both SXR fluxes and their corresponding time derivatives (dashed lines in Figure~\ref{fig1}(a)) manifest slight bumps around 03:40~UT, and significant enhancements around 03:54~UT. We can split the event into the pre-impulsive stage ($\sim$03:36-03:46~UT), the impulsive stage ($\sim$03:46-03:55~UT), and the gradual stage (after 03:55~UT).

\begin{figure}
\centering
\epsscale{1.0}
\plotone{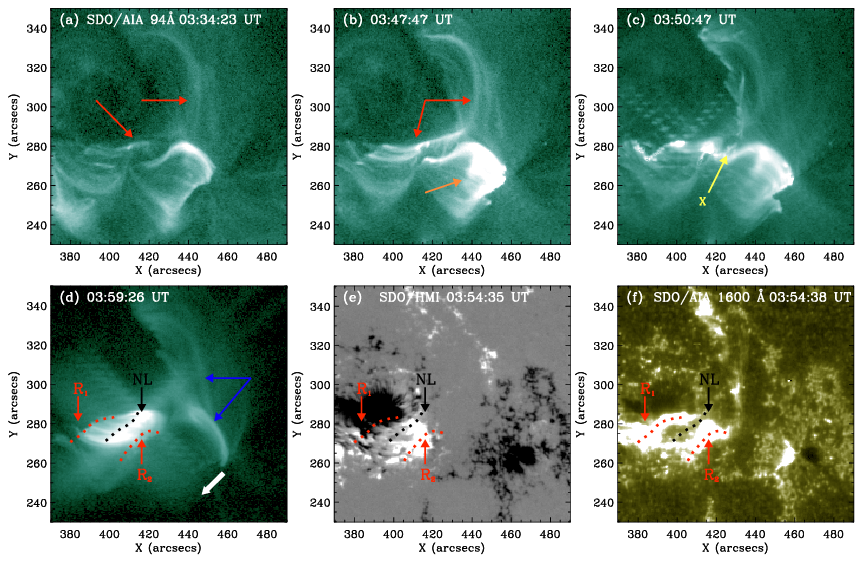}
\caption{The AIA 94~\AA{} images(a - d), the HMI magnetogram (e), and 1600~\AA{} image (f). The red arrows in panel (a) point at the loops to be reconnected in the pre-impulsive stage, and the red ones in panel (b) point at the just-reconnected loop. The symbol $^{``}X^"$ in panel (c) points at the X point of the subsequent reconnection during the impulsive stage. Panel (d) presents the erupting (blue arrows) and post-flare loops. The thick arrow (white) indicates the moving direction of the brightening loop cusps. The two red dashed line in panels (d)-(f) delineate the two UV ribbons ($R_1$ and $R_2$) and the black dashed line delineates the location of the loop cusps ($NL$). An animation for AIA 94~\AA{} (for panels (a-d)) from 03:34:47~UT to 04:01:35~UT showing the dynamic evolution of the flare is available. The real-time duration of the animation is 13s.%, (c), and (d) points at the reconnecting loops ($L3$ \& $L4$), reconnection X point ($X$) and newly reconnected erupting loops ($L5$) as well as post-flare loops ($L6$) in the impulsive stage. $L5$ and $L6$ are overlaid on the 1600~\AA{} image and magnetogram.
}
\label{fig2}
\end{figure}

Figure~\ref{fig2} shows the dynamic evolution with AIA images at 94~\AA{}. During the pre-impulsive stage (Figure~\ref{fig2}(a)), two sets of loops (pointed by arrows) meet and reconnect after $\sim$03:36~UT, generating a large-scale coronal loop system (red arrows in Figure~\ref{fig2}(b)). These new loops rise and approach the southwestern loops (orange arrows), leading to sequential reconnection around the $^{``}X^"$ point (Figure~\ref{fig2}(c)) during the impulsive stage. Then the loops (blue arrow in Figure~\ref{fig2}(d)) erupt, and sequential brightening appears along the neutral line between reconnecting loops ($NL$ in Figure~\ref{fig2}(d)). According to the magnetogram data (Figure~\ref{fig2}(e)), the footpoints of the flare loops have opposite magnetic polarities, being associated with a pair of Ultraviolet (UV 1600~\AA{}) ribbons ($R_1$ \& $R_2$, Figure~\ref{fig2}(f)).

Figure~\ref{fig1} presents the microwave and HXR data. According to the microwave temporal profiles, the flux densities above $\sim$9.4~GHz exceed 1000~SFU, and those above 35~GHz reach up to $\sim3000$~SFU (see Figure~\ref{fig1}(b) and (c)). Six distinct local peaks (black arrows in Figure~\ref{fig1}(c), marked as $P_1-P_6$) can be identified from 03:51 to 03:53~UT and appear in all microwave flux curves. The intervals between these peaks are 22s, 17s, 9s, 16s, and 9s, without significant periodicity. This is why we do not classify this flare as a QPP event. The peaks are more prominent at frequencies higher than 10~GHz. The HXR fluxes above $\sim$100~keV are 1-2 orders of magnitude above the corresponding background values, with similar peaks (Figure~\ref{fig1}(d)).

From Figure~\ref{fig1}(c) the flux densities at 17~GHz (orange dashed) are close to those at 35~GHz (green dashed) at all local peaks. This indicates that the microwave spectral peak/turnover frequency should lie around or exceed 17~GHz. In other words, the current event belongs to flares with a high turnover frequency \citep[c.f.,][]{Yan2023}.

\section{Analysis of the multi-peak radiation}
\subsection{The Main Peak}
\begin{figure}
\centering
\epsscale{0.9}
\plotone{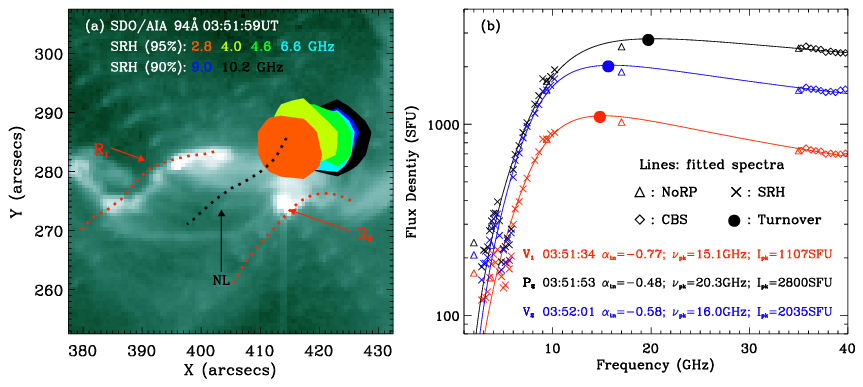}
\caption{Imaging and spectral data around the major peak ($P_2$$\sim$03:51:53~UT) of the M6.5 flare. Panel (a) presents 94~\AA{} image at 03:51:59~UT overlaid with the SRH microwave contours (filled) at 2.8 (orange), 4.0 (yellow), 4.6 (green), 6.6 (cyan), 9.0 (blue), and 10.2 (black)~GHz. The red and black dashed lines delineate the two flare ribbons ($R_1$ and $R_2$) and magnetic field neutral line
% the loop cusp
 ($NL$) of sequential brightening, respectively. Panel (b) presents the fitted spectra of NoRP (triangle), SRH (cross), and CBS (diamond) data at three moments (corresponding to $V_1$, $P_1$, and $V_2$ in Figure~\ref{fig1}(c)). The fitted optically-thin spectral index ($\alpha_{tn}$), turnover frequency ($\nu_{t}$, as filled circle), and turnover flux ($I_{t}$) are given.}
\label{fig3}
\end{figure}

We first pay attention to the highest peak ($P_2$) around 03:51:53~UT. Figure~\ref{fig3}(a) presents the 94~\AA{} image observed at 03:51:59~UT, overlaid with microwave sources (filled contours). The microwave sources appear between the two UV ribbons ($R_1$ and $R_2$), being co-spatial with the top-cusp of top section of the flare loop. Such type of the observed microwave brightness distribution, with the maximum brightness close to the looptop can be explained by effects of localized injection and transverse pitch-angle anisotropy of accelerated high-energy electrons as well as by possible high optical thickness of the microwave source in the looptop where magnetic field strength is minimal \citep[e.g.][]{Melnikov2002,Kuznetsov2015}.

We observe some spatial separation among these microwave sources (see filled contours in Figure~\ref{fig3}(a)). In the lower-frequency range (From $\sim$2.8 to $\sim$6.6~GHz), the sources manifest regular spatial dispersion with frequency, with lower-frequency sources locating to the left and higher-frequency sources locating to the right. The source spacing between 2.8 (orange) and 6.6~GHz (cyan) reaches up to $\sim$10$^{''}$. The higher-frequency sources at 6.6 (cyan), 9.0 (blue), and 10.2~GHz (black) overlap with each other, and are closer to the right ribbon $R_2$. These sources together agree with the overall loop top morphology of the flare. The overlap of higher-frequency sources may stem from the projection effect near $R_2$.

We further perform a spectral fitting \citep[e.g. see function in][]{Ning2007,Asai2013} of the spatially-unresolved nonthermal microwave spectra observed by SRH, NoRP, and CBS. We note that significant error may arise from data discrepancies among instruments due to their different calibration methods. To constrain systematic errors, we cross-calibrate the SRH and CBS data with the NoRP ones measured at the same frequency: we multiply the SRH fluxes with the ratio of NoRP to SRH values at 9.4~GHz, and the CBS fluxes with the ratio of NoRP to CBS values at 35~GHz. %SRH fluxes multiplying the ratio of NoRP value to SRH and CBS values at 9.4 and 35~GHz, respectively. Figure~\ref{fig3}(b) presents the fitted spectra and parameters around the major peak (see Figure~\ref{fig1}(c)).
The fitted results (see Figure~\ref{fig3}(b)) are representative of typical gyrosynchrotron spectra (solid lines) with large turnover frequencies ($\nu_{t}$) of $>15$~GHz (filled circle). At 03:51:34, 03:51:53, and 03:52:01~UT, $\nu_{t}$ is $\sim$15.1, 20.3, and 16.0~GHz, and the turnover (peak) flux $I_{t}$ is $\sim$1107, 2800, and 2035~SFU, respectively. The optically thin spectra are very hard at the three moments, with index $\alpha_{tn}$ being $-0.77$, $-0.48$, and $-0.58$, respectively, presenting an overall soft-hard-soft spectral pattern. %All the three spectral parameters ($\nu_{t}$, $I_{t}$, and $\alpha_{tn}$) correlate positively with the total flux densities.

\subsection{Other Peaks}
\begin{figure}
\centering
\epsscale{1.0}
\plotone{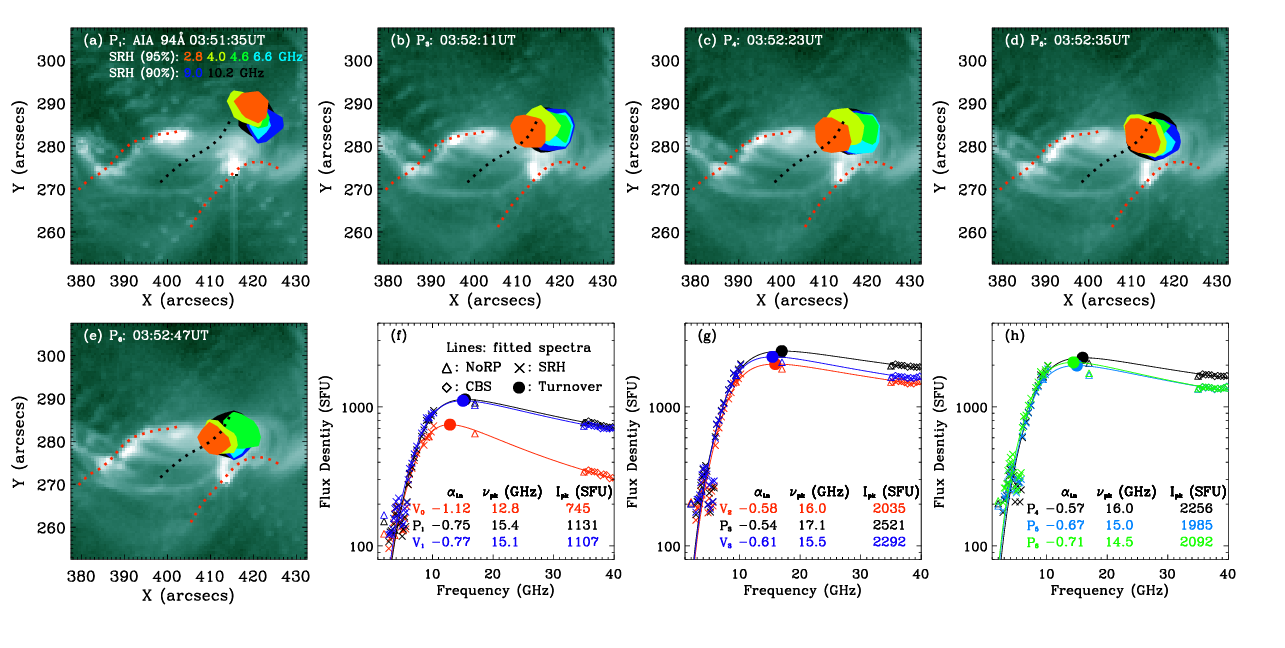}
\caption{Same as Figure~\ref{fig3} but for the other five peaks ($P_1$, $P_3$, $P_4$, $P_5$, and $P_6$). Panels (a)-(e) show EUV images overlaid with the corresponding microwave contours. Panels (f)-(h) show the data and the fitted microwave spectra: (f) for $P_1$, (g) for $P_3$, and (h) for $P_4$, $P_5$, and $P_6$.
}
\label{fig4}
\end{figure}

In Figure~\ref{fig4}(a)-(e), we overplot the microwave contours on EUV images observed around other peaks. Their characteristics are similar to those of the major peak: 1) the microwave sources are between the corresponding flare loop footpoints; 2) sources at lower frequency present spatial dispersion, while sources at frequency above $\sim$6.6~GHz overlap and are closer to $R_2$.

Figure~\ref{fig4}(f)-(h) presents the fitted gyrosynchrotron spectra for 9 selected moments (see arrows in Figure~\ref{fig1}(c)). During the first peak, the fitted spectra reveal that $\nu_{t}$ increases from 12.8~GHz at $V_0$ to 15.4~GHz at $P_1$, and $\alpha_{tn}$ varying from $-1.12$ at $V_0$ to $-0.75$ at $P_1$, with increasing flux density. For other peaks, $\nu_{t}$ (GHz) and $\alpha_{tn}$ are 15.4 and $-0.75$ at $P_1$, 17.1 and $-0.54$ at $P_3$, 16.0 and $-0.57$ at $P_4$, 15.0 and $-0.67$ at $P_5$, and 14.5 and $-0.71$ at $P_6$, respectively. We see that $\alpha_{tn}$ manifests a general decreasing trend from $P_3$ to $P_6$.

We conclude that all these peaks (including the major peak ($P_2$)) present similar properties with high turnover frequency ($\nu_{t}>$15~GHz) and hard optically-thin spectra with $\alpha_{tn}$ being larger than -0.8.

\subsection{Source and spectral evolution}
\label{sec3.3}
\begin{figure}
\centering
\epsscale{1.0}
\plotone{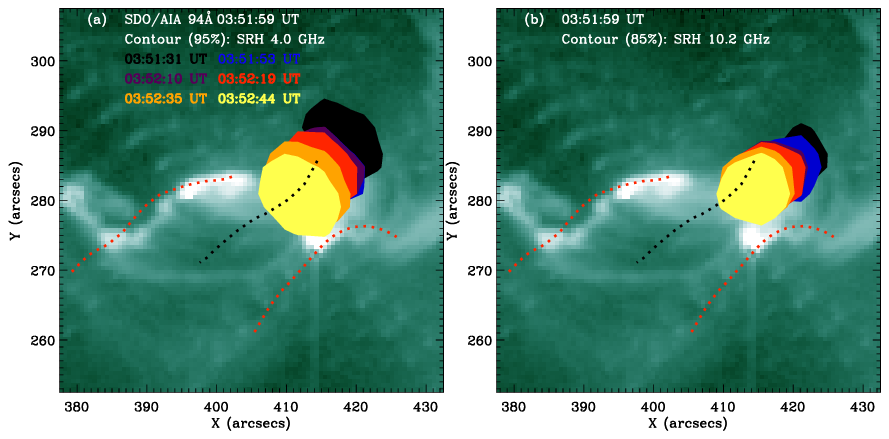}
\caption{Motion of the microwave sources at 4.0~GHz (a) and 10.2~GHz (b) overlaid onto the SDO/AIA 94~\AA{} image (03:51:59~UT). The SRH contours were observed at 03:51:31 ($P_1$, black), 03:51:53 ($P_2$, blue), 03:52:10 ($P_3$, purple), 03:52:19 ($P_4$, red), 03:52:35 ($P_5$, orange), and 03:52:44 ($P_6$, yellow)~UT,.
}
\label{fig5}
\end{figure}

We now analyze the overall evolution of the emission sources during the impulsive stage. Figure~\ref{fig5} shows the microwave contours overplotted on AIA 94 {\AA} images  at 4.0 (a) and 10.2~(b)~GHz, at the six local peaks. We observe systematic source movements from northwest to southeast from 03:51:31~UT to 03:52:44~UT, coinciding with the sequential EUV brightenings along photospheric magnetic field neutral line $NL$ (see Figure~\ref{fig2} and the accompanying animation). The moving distance decreases with increasing frequency. For instance, the source centroid moves $\sim 15-20''$ at 4.0~GHz (see Figure~\ref{fig5}(a)), and $\sim 5-10''$ at 10.2~GHz (Figure~\ref{fig5}(b)). The visible movement of the microwave source from peak to peak possibly originates from the process of new flaring loops appearing sequentially along the NL. A similar process was reported for the flare 22 Aug 2005 \citep{Reznikova2010}

\begin{figure}
\centering
\epsscale{.9}
\plotone{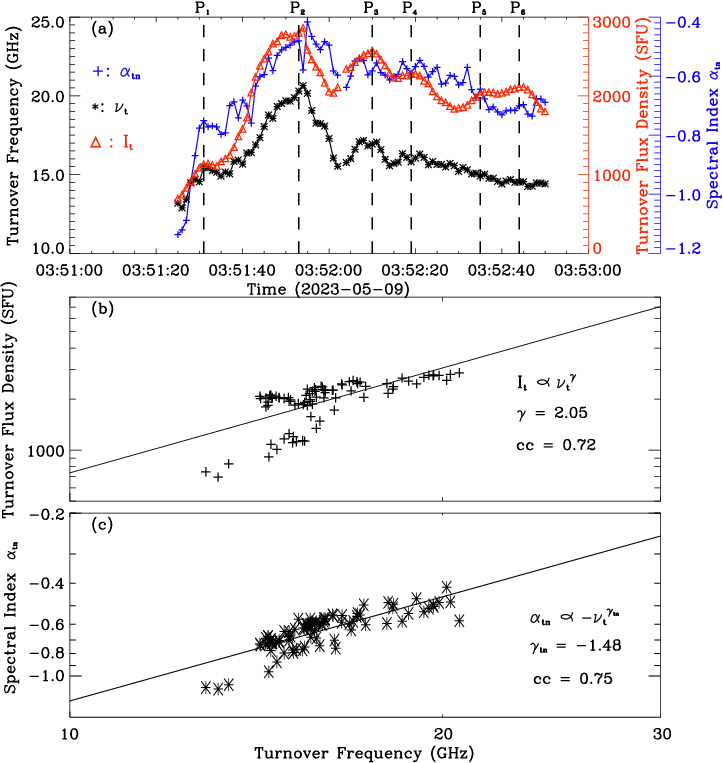}
\caption{Evolution of the microwave spectral parameters from 03:51:00 UT to 03:53:00~UT. Panel (a) presents the fitted optically-thin spectral indices ($\alpha_{tn}$), the turnover frequency and the corresponding intensity ($\nu_{t}$ and $I_{t}$). Panel (b) and (c) presents the fitting of $I_{t}$ and $\alpha_{tn}$ versus $\nu_{t}$, respectively. The dashed line in panels (a) denotes the local peaks.
}
 \label{fig6}
\end{figure}

In Figure~\ref{fig6}, we show the evolution of the fitted spectral parameters ($\alpha_{tn}, \nu_{t}$, and $I_{t}$) from 03:51 to 03:53~UT. During the whole impulsive stage $\nu_{t}$ is close to or above $\sim$15~GHz, and $I_{t}$ exceeds $\sim$2000~SFU. Both parameters correlate well with the total microwave flux densities. According to Figures~\ref{fig6}(b) and (c), both the stronger turnover flux density ($I_{t}$) and harder optically-thin spectral slope ($\alpha_{tn}$) correlate well with the higher corresponding turnover frequency ($\nu_{t}$). Relation between $I_{t}$ and $\nu_{t}$ can be fitted with a power-law dependence of $I_{t}\propto\nu_{t}^{2.05}$, with a correlation coefficient being $\sim 0.72$, and that between $\alpha_{tn}$ and $\nu_{t}$ can be fitted with ${\alpha_{tn}}\propto-\nu_{t}^{-1.48}$, with a correlation coefficient being $\sim 0.75$.

Around $P_1$, the spectral power-law index $\alpha_{tn}$ increases sharply from $\sim -1.1$ to $\sim -0.4$ in about 20~s, reaching maximum around $P_2$ ($\sim$03:51:53~UT). After that, $\alpha_{tn}$ decreases to $\sim -0.7$. The overall evolution of $\alpha_{tn}$ manifests a weak $^{``}$soft-hard-soft$"$ (SHS) trend, with $\alpha_{tn}$ varying from $\sim -1.12$ to $-0.42$. Note that the optically-thin spectra during this peak are very hard, indicating that the energetic electrons have hard energy spectra ($\sim -1.8<\delta<\sim-2.6$), according to the empirical relation of $\delta=(\alpha_{tn}-1.22)/0.9$ given by \citet{Dulk1985}. This means very efficient acceleration of high energy electrons in the flare.

\begin{figure}
\centering
\epsscale{.75}
\plotone{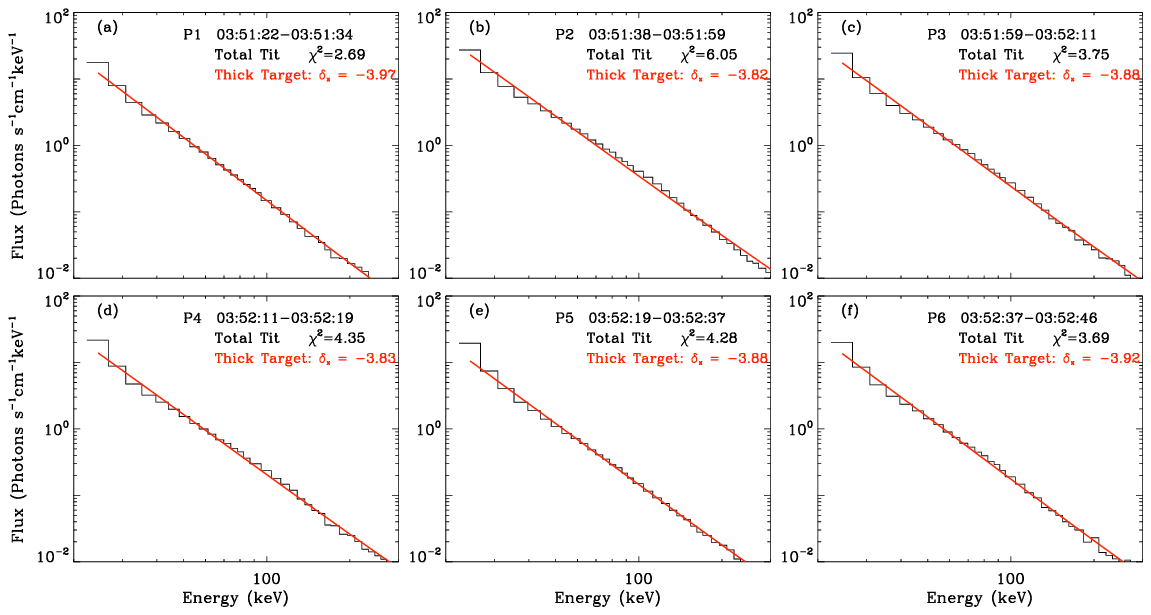}
\caption{The HXR data of KW (black) and the thick-target spectral fitting with a single power law (red) around the flare peaks (black arrows in Figure~\ref{fig1}(c)). The fitted parameters are written.
}
 \label{fig7}
\end{figure}

We also performed the spectral fitting of the KW data (see Figure~\ref{fig7}). This is done with a thick-target model. During the impulsive phase, the electron energy spectral index ($\delta_x$) varies from $-3.97$ to $-3.82$, with the maximum being at the peak $P_2$ and the minimum being at the peak $P_1$. The $\delta_x$ undergoes a trend similar to its microwave counterpart ($\delta$). Both the HXR and microwave spectra get hardened from $P_1$ to $P_2$, and then get softened from $P_2$ to $P_6$. Note that the difference between $\delta_x$ and $\delta$ is $\sim 2$, agreeing with previous studies \citep[reviewed by][]{White2011}.

\section{Modeling microwave emission distribution}
We further modeled the event with the GX simulator \citep{Nita2023}, which is based on the fast gyrosynchrotron code to calculate the microwave emission \citep{Kuznetsov2021}.

\subsection{Basic setting of the flux tube}
\label{fp1}
\begin{figure}
\centering
\epsscale{0.75}
\plotone{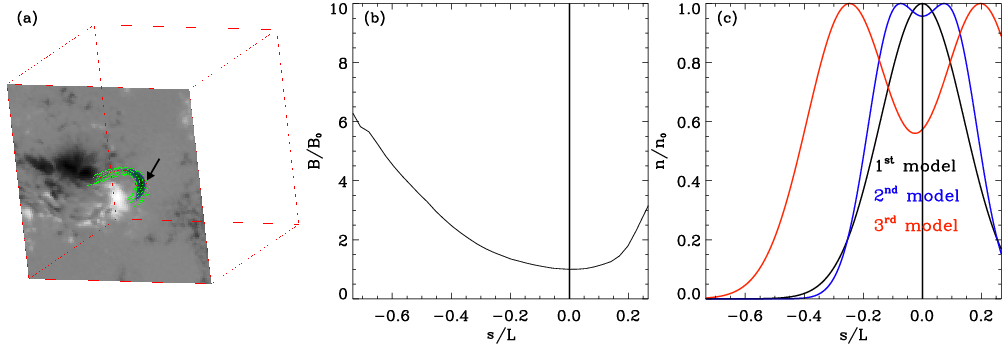}
\caption{Parameter setup of the GX Simulator. Panel (a) presents the selected flux tube extrapolated with NLFFF extrapolation (line-of-sight view), showing the spatial distribution of nonthermal electrons as the shadowed area. Panel (b) presents the normalized magnetic field strength, and panel (c) presents the normalized density of the nonthermal electrons, along the loop. The arrow in panel (a) and vertical lines in panels (b) and (c) denote the location of the loop top.
}
\label{fig8}
\end{figure}

The magnetic configuration of the flare loop (see upper panel of Figure~\ref{fig8}(a)) was deduced via the nonlinear force-free field \citep[NLFFF,][]{Wiegelmann2007} extrapolation of vector magnetogram obtained by the Helioseismic and Magnetic Imager \citep[HMI,][]{Schou2012} at 03:48:00~UT. We selected the central flux tube with the loop length $L\approx2.48\times10^9$~cm, the looptop field strength $B_0=190$~G, and the left and right footpoints field strength $B_{LF}\approx1214$~G and $B_{RF}\approx600$~G.  The loop top is 0.733$L$ from the left footpoint. Figure~\ref{fig8}(b) presents the values of $B$/$B_0$ along the loop. We assumed the thermal electron plasma has a temperature $T=2\times10^7$~K and number density $N=5\times10^9~\rm{cm}^{-3}$, distributed along the loop according to the function prescribed by \citet{Gary2013}.

\subsection{Modeling electron distributions along the flux tube}
\label{fp2}
To understand the origin of the observed microwave brightness distribution with the peak closer to the right footpoint, we do modeling of the spatial distribution of nonthermal electrons along a magnetic loop by solving the kinetic Fokker-Planck equation. Here we consider Fokker-Planck equation, which includes nonstationary continuous injection of particles and take into account Coulomb collisions and magnetic mirroring \citep{Hamilton1990, Reznikova2009}:
$$
\begin{array}{c}
\displaystyle
\frac{\partial f}{\partial t}= -c \beta \mu
\frac{\partial f}{\partial s} + c \beta \frac{d \, ln B}{d s}
\frac{\partial }{\partial \mu}\left[ \frac{1-\mu^2}{2} f \right]
\end{array}
$$
\begin{equation}
 \ \
 + \frac{c}{\lambda _0}\frac{\partial}{\partial E}\left(
\frac{f}{\beta} \right)
+ \frac{c}{\lambda _0 \beta ^3 \gamma ^2}
 \frac{\partial}{\partial \mu} \left[  \left(1- \mu ^2 \right)
 \frac{\partial f}{\partial \mu} \right] + S,
 \label{F_P}
\end{equation}
where $f=f(E,\mu,s,t)$ is electron distribution function of kinetic energy $E=\gamma -1~$  (in units of $mc^2$), pitch-angle cosine $\mu=cos~\alpha$, distance from the flaring loop center $s$, and time $t$, $S=S(E,\mu,s,t)$ is injection rate, $\beta=v/c$, $v$ and $c$ are electron velocity and speed of light, $ \gamma =1/\sqrt{1-\beta^2}$ is Lorentz factor, $B=B(s)$ is magnetic field distribution along the loop, $\lambda_0 =10^{24}/n(s)ln~\Lambda,$ $n(s)$ is plasma density distribution, $ln \Lambda$ is Coulomb logarithm.

For solving the equation, we follow the initial and boundary conditions suggested in \citet{Reznikova2009}. Most of the model parameters are taken to be close to those obtained in Sections 2 and \ref{fp1}. The left part of the model loop with the negative values of s has a stronger magnetic field than the right one, and corresponds to the eastern foot of the observed loop. The injection function $S(E,\mu, s, t)$ is supposed to be a product of functions dependent only on one variable (energy $E$, cosine of pitch angle $\mu$, position $s$, and time $t$):  $S(E,\mu, s, t ) = S_1(E)\times S_2(\mu)\times S_3(s) \times S_4(t)$, where the energy dependence is a power law: $S_1(E) = (E/Emin)^{-\delta}$, $E_{min}$ = 30 keV, with spectral index $\delta$ = 2.6 that is equal to one derived from microwave spectrum; $S_2(\mu)$ is a pitch-angle distribution; $S_3(s)$ is an injection source spatial distribution: $S_3(s) = \rm{exp}(-(s - s_0)^2/s_1^2) $; $S4(t)$ is a time dependence: $S4(t ) = \rm{exp}((t - t_m)^2/t_0^2)$, $t_m = 15$s, $t_0 = 12$s, the half-width of injection duration is 30s, similar to a single peak duration of the flare under study.

\begin{figure}
\centering
\epsscale{0.95}
\plotone{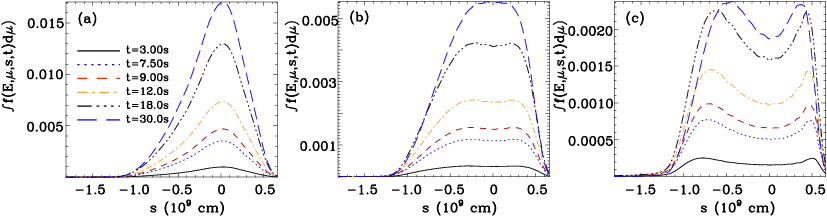}
\caption{The simulated evolution of nonthermal electron distributions along the flux tube for the $1^{st}$ (a), $2^{nd}$ (b), and $3^{rd}$ (c) models. The lines denote the distributions at 3.0, 7.5, 9.0, 12.0, 18.0, and 30.0s after the electron injection.
}
\label{fig8s}
\end{figure}

We have considered three models with the isotropic injection ($S_2(\mu) = 1$): the first one with the injection location at the looptop ($s=0$), the second one with the injection location close to the right footpoint ($s=0.20L$), and the third one with the injection close to the left footpoint ( ($s=-0.40L$)).
The obtained electron number density distributions along the loop for energies of electrons $E=405$~keV at different moments of time are shown in Figure~\ref{fig8s}(a), (b), and (c), respectively.

The first model (Figure~\ref{fig8s}(a)) presents a single peak of the electron distribution at the location with minimum magnetic field strength ($s = 0$) in the rising, peak, and decay phases of the flare. This happens because electrons with large pitch-angles are accumulated in the local magnetic trap mostly around $B(s)=B_{min}$, while electrons with small pitch-angles precipitate into the dense chromosphere. Two spatially separated peaks appear for the $2^{nd}$ and $3^{rd}$ models (Figure~\ref{fig8s}(b) and (c)), with the minimum at the looptop. The reason for this is that energetic electrons with large pitch-angles which originally isotropically injected at the positions in one loop leg have their reflection in the opposite leg where the magnetic field strength is the same \citep[e.g.][]{Kuznetsov2011,Kuznetsov2015}. The shape of the distributions remains more or less similar in case we increase Coulomb scattering by 10 times, increasing the number density of thermal plasma in the looptop up to  $N = 5 \times 10^{10}$ cm$^{-3}$.

%\subsection{Simulation results}
\subsection{Modeling gyrosynchrotron emission distributions}

We simulate microwave emission with GX\_Simulator for the three models with analytical distributions with the shapes similar to the ones obtained from Fokker-Planck simulations:
$$
\begin{array}{c}
f_1(s) = \rm{exp}(-(5s/L)^2),
\end{array}
$$
$$
\begin{array}{c}
f_2(s) = \rm{exp}(-(8(s/L-0.1))^2) + \rm{exp}(-(8(s/L+0.1))^2),
\end{array}
$$
\begin{equation}
f_3(s) = \rm{exp}(-(5(s/L-0.2))^2) + \rm{exp}(-(5(s/L+0.25))^2),
\end{equation}
where $f_1$, $f_2$, and $f_3$ are the normalized spatial distributions of nonthermal electrons along the flare loop for the 1$^{st}$, $2^{nd}$, and $3^{rd}$ models (Figure~\ref{fig8}(c)).

\begin{figure}
\centering
\epsscale{0.95}
\plotone{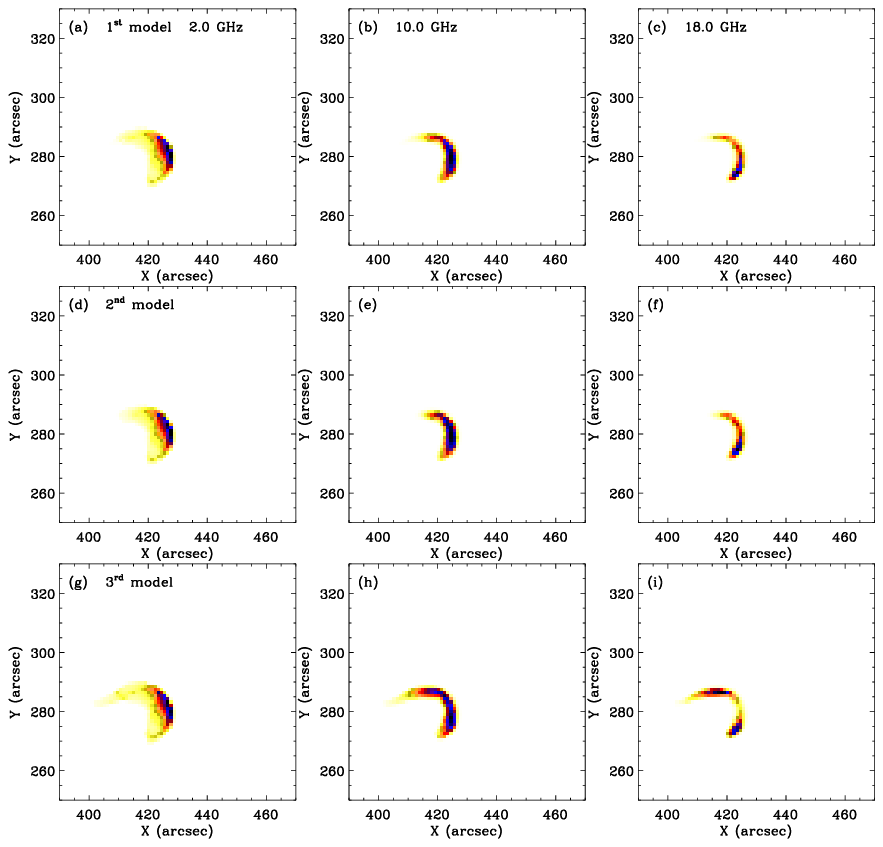}
\caption{The simulated microwave source distributions for three models of nonthermal electrons distribution in Figure~\ref{fig8}(c): (a)-(c) for the 1$^{st}$ model, (d)-(f) for the $2^{nd}$ model, and (g)-(i) for the $3^{rd}$ model. The left, middle and right panels present the modeled images at 2.0, 10.0, and 18.0~GHz, respectively. The maps are not convolved with the instrument beam
}
\label{fig8b}
\end{figure}

We show the simulated microwave source distributions in Figure~\ref{fig8b}. Its comparison with Figure~\ref{fig3}(a) shows that the observed source shift to the right footpoint with frequency agrees with the first two models better than the third one. So, we conclude that the electrons are accelerated and injected close to the top of the observed flaring loop, possibly nearing the cusp of reconnecting field lines, in the flare under study.

We further fix the prescribed electron spatial and energy distributions as those used in the 1$^{st}$ model, but varied the number density $n_0$ from 2.1$\times10^8$, 1.9$\times10^8$, to 1.7$\times10^8~\rm{cm}^{-3}$ for peaks $P_2$, $P_3$, and $P_4$, respectively. %The modeled source position and spectra of microwave emission at different frequencies agree with the observations.

\begin{figure}
\centering
\epsscale{0.95}
\plotone{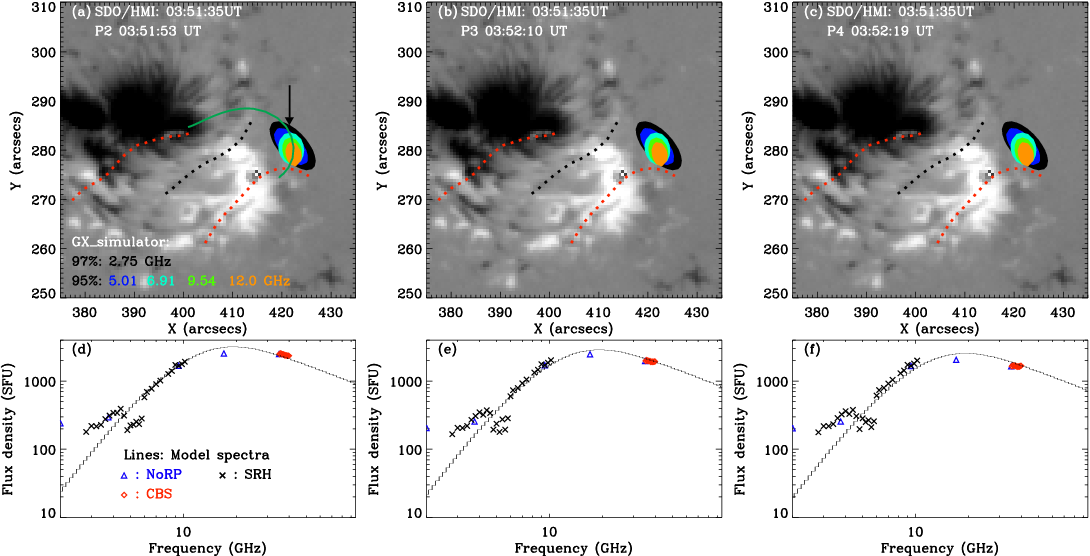}
\caption{Upper panels (a-c): the simulated microwave sources overlaid onto the magnetogram (at 03:51:35~UT) at the 5 selected frequencies (contour level of 97\% for 2.75~GHz, and 95\% for 5.01, 6.91, 9.54 and 12.0~GHz). lower panels (d-f): the simulated spectra compared with the NoRP (triangles), SRH (crosses) and CBS (diamonds) data, for peaks $P_2$, $P_3$, and $P_4$. The simulated sources have been convolved with the SRH beam at 03:52:00~UT. The green line in panel (a) depicts the central line of the modeled flux tube, and arrow points at the location of the loop top (see Figure 8).
}
\label{fig9}
\end{figure}

The modeled sources (upper panels in Figure~\ref{fig9}) concentrate around the loop top (an arrow in  Figure~\ref{fig9}(a)), agreeing with the SRH observations in Figure~\ref{fig3} and \ref{fig4}. The source separations are evident, with low-frequency sources (2.75 (black), 5.01 (blue), and 6.91 (cyan) GHz) nearing the loop top, and high-frequency sources (9.54 (green) and 12.0 (orange) GHz) being closer to the footpoint. The spacing reaches up to $\sim5^{\prime\prime}$ between 2.75 and 12.0 GHz.

The modeled spectra also agree with the observations (lower panels in Figure~\ref{fig9}). They present typical gyrosynchrotron patterns with large $\nu_{t}$ that is $\sim$19, 18, and 17~GHz for $P_2$, $P_3$, and $P_4$, respectively, being close to those obtained above (see Figures~\ref{fig3} and~\ref{fig4}).

\section{Summary}
% and Discussion}
We reported an M6.5 solar flare occurring on 2023 May 9. The flare presents intriguing multi-peak profiles in almost all available flux densities of microwave and HXR. Six local peaks are observed during the impulsive stage, with quite similar characteristics in terms of the microwave spectra and source patterns.

In terms of the microwave images, the SRH sources lie between the two UV footpoint ribbons, corresponding to the looptop/cusp section. For most peaks, the SRH sources at lower frequencies ($<\sim$6.6~GHz) present clear spatial dispersion, while those at higher frequencies overlap with each other. From peak to peak, these sources move from northwest to southeast on the disk over time.
%We consider that this movement originates from new flaring loops which appear sequentially along the NL. This is supported by the observed similar movement of EUV brightenings.

In terms of the microwave spectra, the optically-thin spectral indices ($\alpha_{tn}$) vary from $\sim -1.2$ to $-0.4$, and the spectra are very hard in general with a gentle first-hardening-then-softening trend similar to the hard X-ray spectral evolution. Note, however, that there is a difference between $\delta_x$ and $\delta$ of $\sim 2$, agreeing with previous studies \citep[reviewed by][]{White2011}.
The spectral turnover frequencies ($\nu_{t}$) of all the temporal peaks remain in the range of $\sim 15-22$~GHz. The higher turnover frequency correlates with a larger turnover flux density and a harder optically-thin spectrum, with power-law dependence of $I_{t}\propto\nu_{t}^{2.05}$, and ${\alpha_{tn}}\propto-\nu_{t}^{-1.48}$, respectively. These spectral features indicate strong acceleration of high energetic non-thermal electrons around these peaks.

According to our microwave data and GX simulations, the energy spectral index of the nonthermal electrons lies in a range of -1.8 to -2.6.
Regarding the spectral turnover frequency, both the Razin effect  and the self-absorption effect can affect its value \citep{Razin1960, Ramaty1969}. In our event, the high turnover frequency $\nu_{t}$ and the very-hard optically-thin spectra favours the latter effect since the larger number density of high energy electrons can shift the spectral maximum toward higher frequency, without significantly enhancing their flux density at lower frequencies \citep[see][]{Melnikov2008,Wu2019}. This is also supported by the found  positive correlation between $I_t$ and $\nu_t$ for this event.

The observed microwave features, including spectra, source position and spatial dispersion at different frequencies have been explained using NLFFF extrapolated magnetic flux tube, Fokker-Plank equation solution for electron distribution along the magnetic tube, and GX\_Simulator simulation for gyrosynchrotron emission. We conclude that the particle acceleration in this flare happens in a compact region in the right portion close to the looptop, possibly in the cusp of reconnecting magnetic field lines.

The features of all observed peaks of emission are similar and differ only in the magnitude of the flux density. It seems that all of them originate from similar
flare loops with the same spatial and energy distributions of the nonthermal electrons, but with different electron number density varying in a range of $\sim$30\%. The GS simulations suggest the peaks are generated by nonthermal electrons that concentrate around the right side of the loop-top region.
% \citep[e.g.][]{Kuznetsov2011,Kuznetsov2015}.
The source separation for each peak at different frequencies is consistent with our GS spatial distribution modeling, according to which stronger magnetic field (being at lower altitude) favours generation of higher-frequency emission \citep{Nindos2020}.

With observations and the simulations, we suggest that the observed multiple peaks during the impulsive stage of this flare stem from intermittent energy release and electron acceleration. The visible movement of the microwave source from peak to peak originates from the process of new flaring loops appearance sequentially along the NL. This is supported by the observed similar movement of EUV brightenings.

%The looptop concentration of nonthermal electrons can account for these imaging features. As homogeneous distributed electrons along loops would generate footpoint microwave source at frequencies even in the optically-thick regime \citep{Kuznetsov2011}. Note that the looptop concentration can be explained by effects of localized injection, transverse pitch-angle anisotropy of injected high-energy electrons or enhanced electron loss at the footpoints \citep[e.g.][]{Melnikov2002,Kuznetsov2015}. The observed source separation for each peak is consistent with the prediction of the GS mechanism: for specified electrons, stronger magnetic field favours generation of higher-frequency emission, implying that higher-frequency sources are closer to the footpoints than the lower-frequency ones \citep[e.g.][]{Nindos2020}. The observed microwave features, including spectra, source position and spatial dispersion, can be modeled with the GX simulator using the same flare loop and the same spatial and energy distributions of the nonthermal electrons, but with different electron abundance varying in a relative range of $\sim$30\%. Both observations and modelling suggest that the observed multiple peaks during the impulsive stage of this flare stem from intermittently accelerated electrons via intermittent reconnection.

\acknowledgments
This work is supported by grants of National Natural Science Foundation of China 42127804, 42074203 and 11873036, the National Key R\&D Program of China under grant 2022YFF0503002 (2022YFF0503000), and by the Russian Science Foundation under grant 22-12-00308 (V.M.). A.K., S.A. and R.S acknowledge financial support by the Ministry of Science and Higher Education of the Russian Federation. We appreciate the teams of SRH, CBS, ASO-S, KW, NoRP, SDO, and GOES for their open data use policy. ASO-S mission is supported by the Strategic Priority Research Program on Space Science, the Chinese Academy of Sciences, Grant No. XDA15320000.


\end{document}